\documentclass[aps,prd,nofootinbib,twocolumn,superscriptaddress,preprintnumbers]{revtex4-2}

\newcommand\blfootnote[1]{%
  \begingroup
  \renewcommand\thefootnote{}\footnote{#1}%
  \addtocounter{footnote}{-1}%
  \endgroup
}
\usepackage{ dsfont }
\usepackage{enumerate}
\usepackage{amsmath}
\usepackage{amsfonts} 
\usepackage{amssymb}
\usepackage{graphicx}
\usepackage{hyperref}
\usepackage{tensor}
\usepackage{siunitx}

\def\mk {{\mathcal K}}

\def\mm {{\mathcal M}}

\def\tr {{\tilde{\rho}}}
\def\p {{\partial}}

\def\re   {{\text{Re}}}

\def\Vol  {{\text{Vol}}}
\def\dsZ  {{\mathds{Z}}}

\begin{document}

\preprint{}
\preprint{}
\rightline{IFT-UAM/CSIC-22-35\blfootnote{joan.quirant\emph{at} estudiante.uam.es}}
\title{\Large Non-integer conformal dimensions for type IIA flux vacua}
\author{\large Joan Quirant}
\affiliation{Instituto de F\'{\i}sica Te\'orica UAM-CSIC, Cantoblanco, 28049 Madrid, Spain }

\begin{abstract}
\vspace{0.5cm}
We study the conformal dimensions of the would-be dual operators to the stabilized moduli for two non-supersymmetric DGKT vacua. For the first of them, related to the standard SUSY DGKT vacuum by $G_4=-G_4^{\text{SUSY}}$, we obtain integer conformal dimensions. For the second of them, which has a non-zero harmonic component in $G_2$, we obtain both integers and real numbers.
\vspace{1cm}
\end{abstract}

\maketitle
\section{Introduction}
Scale separation and moduli stabilisation are two indispensable requirements in any EFT constructed from string theory that aims to describe our universe. As of today, some of the most studied models featuring both characteristics are the so-called DGKT vacua \cite{DeWolfe:2005uu, Camara:2005dc}, AdS$_4$ flux compactifications of massive IIA in Calabi-Yau orientifolds.

Under the approach of the swampland program \cite{Vafa:2005ui} -for  reviews see \cite{Brennan:2017rbf, Palti:2019pca,vanBeest:2021lhn,Grana:2021zvf}- these scenarios have received a renewal interest and are  now being scrutinised, specially since it was conjectured in \cite{Lust:2019zwm} that (SUSY) scale separated vacua may lie in the swampland -see \cite{Gautason:2015tig} for previous discussion and  \cite{Buratti:2020kda} for a refinement of the conjecture\footnote{DGKT vacua are compatible with this refinement and also with the proposal of \cite{Gautason:2018gln}}-. In this regard, its 10d uplift  was addressed in \cite{Acharya:2006ne, Saracco:2012wc, McOrist:2012yc, Junghans:2020acz, Marchesano:2020qvg} whereas  more discussion on scale separation in IIA can be found in \cite{Font:2019uva,  Farakos:2020phe, Baines:2020dmu,DeLuca:2021mcj, Cribiori:2021djm, Apers:2022zjx,Emelin:2022cac, Cribiori:2022trc}.
The study of the holographic description of these vacua was initiated recently in \cite{Aharony:2008wz, Conlon:2021cjk, Apers:2022zjx, Apers:2022tfm}. Scale separation has a clear interpretation in terms of parametric gaps in the spectrum of masses of the dual theory,  and it may be clarified if DGKT vacua are consistent by looking at the kind of CFT that they would correspond to. Surprisingly, it was first noticed in \cite{Conlon:2021cjk} for the original toroidal example, and then generalized in \cite{Apers:2022tfm} for any CY orientifold, that the conformal dimension for the low-lying scalar primaries in the dual CFT of SUSY DGKT vacua is always an integer, independently of the details of the compactification. In the same spirit, \cite{Apers:2022zjx} found that this is also true for the toroidal non-SUSY DGKT models examined there\footnote{But not for the duals of AdS$_3$ DGKT-type vacua and for the example studied in \cite{Font:2019uva} where there is no scale separation.}.

In this letter, we will continue in this direction by examining two of the non-SUSY branches of general (no specific CY orientifold is assumed) DGKT vacua derived in \cite{Marchesano:2019hfb}. As we will show below, for the non-SUSY branch related to the SUSY one by changing $G_4^{ \text{non-SUSY}}= -G_4^{\text{SUSY}}$, the conformal dimension of the scalars dual to the stabilised moduli are  integers as well. Nevertheless, for the so-called branch A2-S1 in that paper, characterised for having a harmonic component in $G_2$ different from zero, $G_2^{\text{harmonic}}\neq 0$, these same conformal dimensions are no longer integers. 

Before presenting the results, it is worth pointing out that non-SUSY AdS vacua of this kind are conjectured to be unstable \cite{Ooguri:2016pdq}. Regarding the status of the two branches at hand, its perturbative stability was verified in \cite{Narayan:2010em,Marchesano:2019hfb}. In addition to this, the non-perturbative stability of the first one ($ -G_4^{\text{SUSY}}$) was first studied in \cite{Narayan:2010em}, where decays were found to be at best marginal, and then in \cite{Marchesano:2021ycx}, where a D8 domain wall instability was found if space-time filling D6s are used to cancel the tadpole.

\section{DGKT vacua in a nutshell}

Consider massive type IIA string theory compactified on an orientifold of $\mathds{R}^{1,3}\times\mathcal{M}_6$ with $\mathcal{M}_6$ a compact Calabi-Yau threefold. Dimensional reduction leads to a $\mathcal{N}=1$  4d supergravity theory whose massless scalar field content is organised as follows \cite{Grimm:2004ua}. On the one hand, there are the complexified K\"ahler moduli,  coming from integrating the K\"ahler 2-form $J=t^a\omega_a$ and the $B=b^a\omega_a$ field
\begin{align}
T^ a=b^a+it^a\, ,	\quad	a\in\left\{1,\dots, h^{1,1}_-\right\}\, ,
\end{align}
where $l_s^{-2}\omega_a$ are  harmonic representatives of  $H_-^ 2\left(\mathcal{M}_6,\left(\mathds{Z}\right)\right)$ and $l_s$ the string length. The metric appearing in the kinetic terms of these moduli is obtained from the K\"alher potential
\begin{align}
K_K=-\log\left(\frac{4}{3}\mk_{abc}t^at^bt^c\right)=-\log\left(\frac{4}{3}\mk\right)\, ,
\end{align}
with $\mk_{abc}=l_s^{-6}\int\omega_a\wedge\omega_b\wedge\omega_b$ and $\mk=6\Vol_{\mm_6}$. On the other hand, there are the complex structure moduli, coming from the complex 3-form $\Omega$, the axio-dilaton and the RR 3-form potential $C_3$. Introducing $\Omega_c\equiv C_3+i\re\left(\mathcal{C}\Omega\right)$ where $\mathcal{C}=e^{-\phi}e^{\frac{1}{2}\left(K_{cs}-K_K\right)}$ is a compensator, with $K_{cs}=-\log\left(-il_s^{-6}\int\Omega\wedge\bar\Omega\right)$ and $\phi$ the 10d dilaton, the complex structure moduli are defined as
\begin{align}
U^\mu=\xi^\mu+i u^\mu=l_s^{-3}\int\Omega_c\wedge \beta^\mu\, ,	\quad	\mu\in\left\{0,\dots, h^{2,1}\right\}\, ,
\end{align}
where we are taking a symplectic basis $\beta^\mu\in H_-^3\left(\mm_6,\mathds{Z}\right)$. Finally, the metric appearing in the kinetic terms of this sector is constructed from the following K\"ahler potential
\begin{align}
K_Q=4\log{\left(\frac{e^\phi}{\sqrt{\Vol_{\mm_6}}}\right)}\equiv -\log\left(e^{-4D}\right)\, .
\end{align}

On top of this background one can add RR and NSNS background fluxes. Following the conventions of \cite{Marchesano:2019hfb} their flux quanta are
\begin{align}
l_s\bar{G}_0&=-m\, ,	& \frac{1}{l_s}\int_{\tilde{\pi}^a}\bar{G}_2&=m^ a\, ,	&	\frac{1}{l_s^3}\int_{\pi_a}\bar{G}_4&=-e_a\, ,	\nonumber \\	\frac{1}{l_s^5}\int_{\mm_6}\tilde{G}_6&=e_0\, ,	&	\frac{1}{l_s^2}\int_{B_\mu} \bar{H}&=h_\mu	\, ,
\end{align}
with $\left[\pi_a\right]\in H_4^+\left(\mm_6,\dsZ\right)$ Poincar\'e dual to $\left[l_s^{-2}\omega_a\right]$ and  $\left[\tilde{\pi}^a\right]\in H^{2}_-\left(\mm_6,\dsZ\right)$ Poincar\'e dual to $\left[l_s^{-4}\tilde\omega_a\right]$, where $l_s^{-6}\int_{X_6}\omega^ a\wedge\tilde{\omega}^b=\delta^b_a$. $B_\mu$ is the 3-cycle de Rham dual to $\beta^\mu$. In the 4d action the presence of fluxes is encoded through the superpotential \begin{align}
l_sW=&e_0+e_a T^a+\frac{1}{2}\mk_{abc}m^aT^bT^c+\frac{m}{6}\mk_{abc}T^aT^bT^c
\nonumber\\&+h_\mu U^\mu\, ,
\end{align}
which involves both the K\"ahler and the complex structure moduli. As shown in \cite{Herraez:2018vae}, the F-term scalar potential generated by this superpotential exhibits a remarkable factorisation between the saxionic and the axionic components. Namely, the potential can be written in full generality as
\begin{align}
V=\frac{1}{\kappa_4^2}\vec{\rho}^{\ t}\, \textbf{Z}\, \vec{\rho}\, ,
\end{align}
where the vector $\vec{\rho}$ depends only on the flux quanta and the axions $\{b,\xi\}$:
\begin{align}
&l_s\rho_0	&	&=	&	&e_0+e_ab^a+\frac{1}{2}\mk_{abc}m^ab^bb^c+\frac{m}{6}\mk_{abc}b^ab^bb^c\nonumber\\ &	&	&	&	&+h_\mu\xi^\mu\, ,\nonumber\\
&l_s\rho_a	&	&=	&	&e_a+\mk_{abc}m^bb^c+\frac{m}{2}\mk_{abc}b^bb^c\, ,\nonumber\\
&l_s\tilde{\rho}^a	&	&=	&	&m^a+mb^a\, ,\nonumber\\
&l_s\tilde{\rho} 	&	&=	&	&m\, ,\nonumber\\
&l_s\hat{\rho}_\mu	&	&=	&	&h_\mu\, ;
\end{align}
whereas the matrix $\textbf{Z}$ depends only on the saxions $\{t,u\}$:
\begin{align}
\textbf{Z}=e^K\left(\begin{matrix}
4	&	&	&	& \\
	&K^{ab}	&	&	& \\
	&	&\frac{4}{9}\mk^2K_{ab}	&	& \\
		&	&	&	\frac{1}{9}\mk^2&\frac{2}{3}\mk u^\mu \\
			&	&	&	\frac{2}{3}\mk u^\mu&K^{\mu\nu} \\
\end{matrix}\right)\, ,
\end{align}
with $K=K_K+K_Q$, $K_{ab}=\frac{1}{4}\partial_{t^a}\partial_{t^b}K_K$ and $K_{\mu\nu}=\frac{1}{4}\partial_{u^\mu}\partial_{u^\nu}K_Q$. This factorisation is maintained even when D6-brane moduli \cite{Escobar:2018tiu} and  $\alpha'$ corrections \cite{Escobar:2018rna}  are included.

Though we will use the $\vec\rho$ language in this note, it may be useful to recall how their components are related with the more familiar gauge invariant field strengths \cite{Herraez:2018vae}. Using the democratic formulation and calling $\textbf{C}\equiv C_1+C_3+C_5+C_7+C_9$  and $
\textbf{G}\equiv d\textbf{C}-H\wedge \textbf{C}+\bar{\textbf{G}}\wedge e^B$, where $\bar{\textbf{G}}$ is just the sum of the previously introduced flux quanta, then
\begin{align}
\rho_0&=\int_{\mm_6} G_6\, ,	&	\rho_a&=\int_{\pi_a} G_4\, ,	&	\tilde{\rho}^a=&\int_{\tilde{\pi}^a} G_2\, ,\nonumber \\\tilde{\rho}&= G_0\, ,	&	\hat{\rho}_\mu&=\int_{B_\mu} H_3\, .
\end{align}

\section{Non-susy vacua and conformal dimensions}

Writing the potential as a bilinear expression is a very useful tool in the search for vacua, as it was exploited in \cite{Escobar:2018tiu, Escobar:2018rna} and especially in \cite{Marchesano:2019hfb}, where a systematic search of extrema for this setup was performed. Among the different branches of vacua found in that paper we will focus on two cases:
\begin{enumerate}[A)]
\item  Branch Non-SUSY$_{G_4}$, characterised by   $\rho_a=-\rho_a^ {\text{SUSY}}$. In this vacuum 
\begin{align}
\hat{\rho}_\mu&=\frac{1}{15}\tilde{\rho}\mk\partial_{u^\mu}K\, ,	&	\rho_a&=-\frac{3}{10}\tilde{\rho}\mk_a\, ,\nonumber\\\nonumber\\ \tilde{\rho}^a&=0\, ,	&	\rho_0&=0\, ,\\\nonumber\\V&=-\frac{4e^K}{75}\mk^2\tr^2\, ; \nonumber\\\nonumber
\end{align}
\item Branch Non-SUSY$_{G_2}$, characterised by  $\tr^a\neq0$. In this vacuum
\begin{align}
\hat{\rho}_\mu&=\frac{1}{12}\tilde{\rho}\mk\partial_{u^\mu}K\, ,	&	\rho_a&=-\frac{1}{4}\tilde{\rho}\mk_a\, ,\nonumber\\\nonumber\\ \tilde{\rho}^a&=\pm\frac{1}{2}\tr\ t^a\, ,	&	\rho_0&=0\, ,\\\nonumber\\ V&=-\frac{e^K}{18}\mk^2\tr^2\, ; \nonumber\\\nonumber
\end{align}
\end{enumerate}
with $\mk_a=\mk_{abc}t^bt^c$. In addition to this, the SUSY branch was studied recently in \cite{Apers:2022tfm}, agreeing with what we obtain. We will not discuss it here.

\subsection{Branch Non-SUSY$_{G_4}$}

All the necessary features of this solution, including the  physical mass of the stabilized moduli, were calculated in  \cite{Marchesano:2019hfb}, namely in  appendix B. We can limit ourselves to  compute the conformal dimension $\Delta\left(\Delta-d\right)=m^2R_{AdS}^2$ of the correspondent fields in the would-be  CFT$_3$ dual. Regarding the saxions of the compactification, its conformal dimension would be
\begin{align}
\label{g41}
\Delta&=10\, ,	&	\Delta_i&=1 \text{ or } 2\, , 	&	\Delta_a&=6\, ,
\end{align}
with $i=1,\dots, h^{2,1}$  and $a=1,\dots,  h_{-}^{1,1}$  taking into account that some fields acquire the same mass. Notice that, as expected, these same results were obtained in \cite{Apers:2022tfm} for the saxionic spectrum of the supersymmetric case, since the mass matrix in both cases is block diagonal and shares the entries of this part. The difference comes from the axions, whose conformal dimensions in the dual theory would be 
\begin{align}
\label{g42}
\Delta&=1\text{ or }2 \, ,	&	\Delta_i&=3\, , 	&	\Delta_a&=8\, ,
\end{align}
with again $i=1,\dots, h^{2,1}$  and $a=1,\dots, h_{-}^{1,1}$. The $\Delta_i=3$ comes from the  fact that only a linear combination of axions appear in the superpotential, so  $ h^{2,1}$ of them remain massless. These results agree and generalize the work of \cite{Apers:2022zjx}, who looked at this kind of vacua but only in toroidal examples. 

\subsection{ Branch Non-SUSY$_{G_2}$}

Unlike the previous case, the mass acquired  by the stabilized moduli was not explicitly computed in the original reference, so some intermedia steps have to be done. We relegate the diagonalization of the Hessian and all the details needed to  appendix \ref{app}. Using again  the relation $\Delta\left(\Delta-d\right)=m^2R_{AdS}^2$,  the conformal dimension of the dual operators for this branch should be 
\begin{align}
\label{g2}
\Delta&=\frac{1}{2} \left(3+\sqrt{393}\right) \, ,	&	\Delta_a&=\frac{1}{2}
   \left(3+\sqrt{201}\right) \, ,	&	\Delta_i&=3 \, ,	\nonumber\\
\Delta&=\frac{1}{2}
   \left(3+\sqrt{33}\right) \, ,	&	\Delta_a&= 6\, ,	&	\Delta_i&=3 \, ,
\end{align}
with again $i=1,\dots, h^{2,1}$ and $a=1,\dots,  h_{-}^{1,1}$. In this case there are $2h^{2,1}$ fields with conformal dimension $\Delta=3$, since the saxionic partners of the usual massless axions do not acquire a mass\footnote{They develop a quartic potential}.

\section{Conclusions}

In this work we have computed the conformal dimension of the low-lying operators of the putative CFT$_3$ dual of two different non-supersymmetric DGKT vacua.

In the first place, we focused on what we called the Non-SUSY$_{G_4}$ vacuum, which has the property of being related to the SUSY one by $G_4=-G_4^{\text{SUSY}}$. We obtained that these dimensions are always integers -see expressions \eqref{g41} for the saxionic sector and \eqref{g42} for the axionic sector-, and totally independent of the details of the compactification. With respect to the saxions this is not new, since this part of the mass matrix is shared with the SUSY branch and the same numbers were obtained in \cite{Apers:2022tfm}. For the axions, this result extends the work of \cite{Apers:2022zjx} -who only looked at toroidal models- to any CY orientifold.

In the second place, we studied what we named the Non-SUSY$_{G_2}$ vacuum, called in this way by having  $G_2^{\text{harmonic}}\neq 0$. Unlike the SUSY and the Non-SUSY$_{G_4}$ vacua, the conformal dimensions of the operators of the dual theory would not be only integers, see \eqref{g2}. The result is again quite simple and does not depend on the details of the compactification. The 10d uplift of this branch has not been studied in detail and it could happen that  problems arise when looked at from the 10d point of view. Following the analysis of \cite{Junghans:2020acz}, there should not be any obstructions in constructing first, a smearing uplift, and then, expanding the solution and localising the source at first order. We have already checked explicitly that indeed the smearing uplift exists. These and more details are being studied in an upcoming work \cite{nuevo}. 

Non-supersymmetric vacua of the kind studied here are conjectured to be unstable \cite{Ooguri:2016pdq}. This could imply that it would not make much sense to study their CFT duals, since these theories could be sick or ill-defined. Indeed in \cite{Marchesano:2021ycx} we showed that for the Non-SUSY$_{G_4}$ branch there seems to be an instability if space-time filling  D6s are used to cancel the tadpole. Opposite to this reasoning, studying these non-supersymmetric vacua from their putative CFT dual could be useful to show that they are unstable, which makes this analysis interesting \textit{per se}.

In fact, constructing the would-be CFT duals of scale separated AdS vacua could be a way of understanding if scenarios of this type are consistent or if they lie in the swampland. In this note, we have shown that not only integers appear when we study the conformal dimensions of the low-lying operators of  DGKT vacua. But we still do not understand why they do appear for the SUSY case or even if the conformal theories we are trying to construct really exist. We hope that both questions could be answered in the not too distant future.

\vspace{10px}
{\bf Acknowledgements}

We would like to thank M. Sasieta for very helpful discussions and explanations and F. Marchesano for very useful discussions and comments on the manuscript. This work is partially supported by the Spanish Research Agency (Agencia Estatal de Investigación) through the Grant IFT Centro de Excelencia Severo Ochoa No CEX2020-001007-S and PGC2018-095976-B-C21, funded by MCIN/AEI/10.13039/501100011033  and by ERDF A way of making Europe. The author is supported through the FPU grant No. FPU17/04293. 

\appendix
\section{Mass spectrum for the non-SUSY$_{G_2}$ vacua}\label{app}

The Hessian for this branch was obtained explicitly in appendix B of \cite{Marchesano:2019hfb}, see equation (B.6). To compute the physical mass spectrum one has to express the fields on a canonical basis. As explained there, for this we decompose the K\"ahler metrics for the K\"ahler and complex structure fields as 
\begin{align}
K_{ab}&=\frac{3}{2\mathcal{K}}\left(\frac{3\mathcal{K}_a\mathcal{K}_b}{2\mathcal{K}}-\mathcal{K}_{ab}\right)\nonumber\\&=\frac{3}{4}\frac{\mk_a\mk_b}{\mk^2} + \frac{3}{2\mk}\left(\frac{\mk_a\mk_b}{\mk}-\mk_{ab}\right)=K_{ab}^{\rm NP}+K_{ab}^{\rm P}\, ,\\
K_{\mu\nu}&=\frac{1}{16}\frac{\p_\mu G\p_\nu G}{G^2} + \frac{1}{4}\left(\frac{3}{4}\frac{\p_\mu G\p_\nu G}{G^2}-\frac{\p_\mu\p_\nu G}{G}\right)\nonumber\\&=K_{\mu\nu}^{\rm NP}+K_{\mu\nu}^{\rm P}\, ,
\end{align}
with $\mk_{ab}=\mk_{abc}t^c$, $K=K_K+K_Q=-\log{\left(\mathcal{G}\right)}$ and $\partial_\mu\equiv\partial_{u^\mu}$. Here $K^{\rm P}$ and $K^{\rm NP}$ refers to the primitive and non-primitive parts of the metric, which act on orthogonal subspaces. One can then express the fields in the canonically normalised basis
\begin{align}
\label{canon}
    ( \xi^\mu \quad b^a \quad u^\mu \quad t^a)  \rightarrow   (\hat{\xi}\quad \hat{b} \quad \xi^{\hat{\mu}} \quad b^{\hat{a}} \quad \hat{u}\quad \hat{t} \quad u^{\hat{\mu}} \quad t^{\hat{a}} )\, ,
\end{align}
where $\hat{\xi}\left( \hat{b}\right)$ is the vector along the subspace corresponding to $K_{\mu\nu}^{\rm NP}\rvert_{\rm vac}\left(K_{ab}^{\rm NP}\rvert_{\rm vac}\right)$, with unit norm, and  $\xi^{\hat{\mu}}\left(b^{\hat{a}}\right)$  correspond to vectors of unit norm with respect to  $K_{\mu\nu}^{\rm P}\rvert_{\rm vac}\left(K_{ab}^{\rm P}\rvert_{\rm vac}\right)$. Analogously, $\left\{\hat{u}\quad \hat{t} \quad u^{\hat{\mu}} \quad t^{\hat{a}} \right\}$ are defined in the same way. In this canonically normalised basis the Hessian reads
\begin{align}
\label{ham}
\textbf{H}=
\mk^2 e^K\tr^2\left(\begin{matrix}
 \frac{8}{9}&\frac{4}{3\sqrt{3}}& 0& 0&0&0&0&0\\
  \\
\frac{4}{3\sqrt{3}}& \frac{14}{9}& 0& 0&0& \pm\frac{8}{9} &0&0\\
\\
0& 0& 0& 0&0&0&0&0\\
\\
0& 0& 0& \frac{14}{9}&0&0&0&\mp\frac{4}{9} \\
\\
0& 0& 0& 0& \frac{8}{9}&-\frac{4}{3\sqrt{3}}&0&0\\
\\
0& \pm\frac{8}{9} & 0& 0&-\frac{4}{3\sqrt{3}}& \frac{26}{9}&0&0\\
\\
0& 0& 0& 0&0&0&0&0\\
\\
0& 0& 0& \mp\frac{4}{9}&0&0&0&\frac{8}{9}\\
 \end{matrix}\right)\, ,
\end{align}
and the physical masses can be obtained  straightforwardly by diagonalising this matrix and dividing by two. Expressed in terms of  $R_{AdS}^2=\frac{3}{|\Lambda|}=\frac{54}{e^K\mk^2\tr^2}$ they are
\begin{align}
m^2=R_{AdS}^{-2}\left\{96\, , \quad  6\, , \quad    48\, ,  \quad   18\, , \quad     0\right\}\, ,
\end{align}
with multiplicity $\left\{1\, ,\quad  1\, , \quad  h^{1,1}_-\, ,\quad h^{1,1}_-\, , \quad 2h^{2,1}\right\}$ respectively.

\bibliography{papers}

\end{document}